\documentclass[prb,aps,preprint,onecolumn,floatfix]{revtex4}
\usepackage{graphics}
\begin{document}
\title{Persistent Currents in Helical Structures}
\author{Menderes Iskin$^1$\footnote{e-mail: menderes.iskin@gonzo.physics.gatech.edu}, I. O. Kulik$^2$}
\affiliation{$^1$School of Physics, Georgia Institute of Technology, Atlanta, GA 30332 \\ $^2$Department of Physics, Bilkent University, Ankara 06533, Turkey}
\date{\today}

\begin{abstract}
Recent discovery of mesoscopic electronic structures, in particular the carbon nanotubes, made necessary an investigation of what effect may helical symmetry of the conductor (metal or semiconductor) have on the persistent current oscillations. 
We investigate persistent currents in helical structures which are non-decaying in time, not requiring a voltage bias, dissipationless stationary flow of electrons in a normal-metallic or semiconducting cylinder or circular wire of mesoscopic dimension.
In the presence of magnetic flux along the toroidal structure,  helical symmetry couples circular and longitudinal currents to each other.
Our calculations suggest that circular persistent currents in these structures have two components with periods $\Phi_0$ and $\Phi_0/s$ ($s$ is an integer specific to any geometry).
However, resultant circular persistent current oscillations have $\Phi_0$ period.
\pacs{PACS:}PACS:73.23.-b
\end{abstract}
\maketitle

Aharonov and Bohm showed that, contrary to the conclusion of classical electrodynamics, there exists effects of the potentials on the charged particles even in the region where all fields vanish. 
This effect has quantum mechanical origin because it comes from the interference phenomenon. 
The well-known manifestation of the Ahoronov-Bohm (AB) effect is the oscillation of electrical resistance and the periodic persistent currents in the normal metal rings and mesoscopic rings threaded by a magnetic flux.
This current arises due to the boundary conditions imposed by the doubly connected nature of the loop.
Therefore, electronic wave function and then any physical property of the ring is a periodic function of the magnetic flux with a fundamental period $\Phi_0$.
In particular, flux dependence of the free energy implies the existence of a thermodynamics (persistent) current.

\emph{Persistent currents in mesoscopic rings:}
Persistent currents in mesoscopic systems was first predicted by one of the authors \cite{c1} and later discovered by Buttiker \cite{c2} et al. 
A number of key experiments also confirmed the existence of persistent currents in isolated rings \cite{c6,c7}.
In the presence of magnetic flux ($\Phi$) applied at the center of the ring, we consider a one dimensional ring of circumference $L_r=2\pi r=N\Delta$.
Here $N$ is the number of lattice points and $\Delta$ is the lattice spacing.
Tight-binding Hamiltonian reads
\begin{equation}
H_e=-t_0\sum_{n=1}^N\left(a_n^+a_{n+1}e^{i\alpha}+h.c.\right) ,\label{ringH}
\end{equation}
where $t_0$ is the hopping amplitude between the nearest-neighbour sites for an undistorted lattice and operators $a_n$ ($a_n^+$) annihilates (creates) an electron at site n.
$\alpha$ is the corresponding phase change which can be expressed in terms of AB flux ($\Phi$)
\begin{equation}
\alpha=\frac{e}{\hbar c}\int_{n}^{n+1}\mathbf{A}\cdot d\mathbf{l}=2\pi\frac{\Phi}{N\Phi_0} ,\label{ringalpha}
\end{equation}
where $\Phi_0=hc/e\simeq4.1\times10^{-7}$ G $cm^2$ is the flux quantum.

Corresponding eigenvalue spectrum and the persistent currents (variation of free energy with the magnetic flux) are both periodic in $\Phi$ with a period $\Phi_0$.
If we ignore spin of the electron, ground state energy and the total current flowing along the ring can be written as
\begin{eqnarray}
E(\Phi)=\sum_n\epsilon_n(\Phi)
=-2t_0\sum_n\cos{\left [\frac{2\pi}{N}\left(n+\frac{\Phi}{\Phi_0}\right)\right]} ,\label{ringE}
\end{eqnarray}
\begin{eqnarray}
I(\Phi)&=&\sum_nI_n(\Phi)=-c\sum_n\frac{d\epsilon_n(\Phi)}{d\Phi} \nonumber \\
&=&-I_0\sum_n\sin{\left[\frac{2\pi}{N}\left(n+\frac{\Phi}{\Phi_0}\right)\right]} ,\label{ringI}
\end{eqnarray}
where $I_0=4c\pi t_0/N\Phi_0$ is the current amplitude, and the summation is over number of electrons, $N_e$, for each value of flux.
 
\emph{Persistent Currents in Helical Structures:}
Persistent currents was believed to be a specific property of isolated systems for a long time \cite{c2}.
However, theoretical studies suggest that persistent currents should also exist in connected rings \cite{c8}.
In the presence of both longitudinal and transverse flux, existence of transverse persistent currents in doubly connected mesoscopic rings is known \cite{c3,c4}.
Moreover, transverse currents may contribute to an experimental observation of longitudinal persistent current and it can substantially increase the amplitude of the AB oscillations \cite{c6}. 

\begin{figure}  [ht]
\centerline{\scalebox{0.4}{\includegraphics{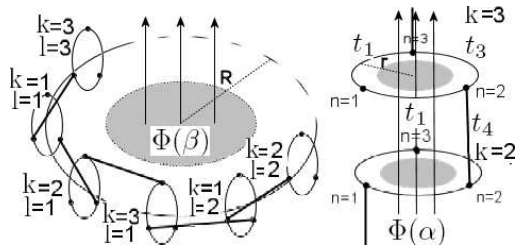}}}
\caption{\label{helical1}
         Left: Toroid of $LN=3*3=9$ rings are connected by vertical hoppings and $\Phi(\beta)$ is applied at the center. 
         Right: Each ring has $N=3$ sites and $\Phi(\alpha)$ is applied at the center.}
\end{figure}

AB effect is also shown to be present in toroidal systems \cite{c5,c9}.
In this paper, we consider a set of identical and connected mesoscopic rings with a circumference of $L_r=2\pi r$ and each having $N$ lattice sites with a lattice spacing of $\Delta_1=L_r/N$. 
We also assume our helical structure has $L$ periods with a periodicity of $N$ rings as shown in Fig. \ref{helical1}.
So, we have a toroid of circumference $L_t=2\pi R$ and it contains $LN$ rings which are uniformly separated by $\Delta_2=L_t/LN$ along the circumference of the toroid.
In the presence of magnetic fluxes $\Phi(\alpha)$ and $\Phi(\beta)$, which are applied through the center of the connected rings and at the center of the toroid respectively, we propose two models.

\emph{First Helical Model:}
In order to have helical symmetry, we allow only nearest-neighbour circular hopping between the sites in each ring and vertical hoppings (no cross-hoppings) between the nearest-neighbour rings along the toroid respectively. 
Assuming the tight-binding model for electron transport, system Hamiltonian can be written as
\begin{eqnarray}
H&=&\sum_{l=1}^L\sum_{k=1}^{N}\sum_{n=1}^{N}\left[-t_1a^+_{lkn}a_{lk,n_N+1}e^{i\alpha} \right. \nonumber \\
&-&(t_3-t_1)\left. a^+_{lkn}a_{lk,n_N+1}e^{i\alpha}\delta_{k,n_N+1} \right. \nonumber \\
&-&(t_4-t_2)\left. a^+_{lkn}a_{l+\delta_{kN}(l_L-l+1),k_N+1,n}e^{i\beta}\delta_{kn} \right. \nonumber \\
&-&t_2\left. a^+_{lkn}a_{l+\delta_{kN}(l_L-l+1),k_N+1,n}e^{i\beta} + h.c.\right] 
\end{eqnarray}
where $t_1$ is the hopping amplitude between the nearest-neighbour sites in the ring,  $t_2$ is the vertical hopping amplitude between the nearest-neighbour rings, $t_4$ is the special hopping amplitude in vertical direction and $t_3$ is the special circular hopping amplitude which connects two special vertical hoppings as shown in Fig. \ref{helical1}.
Note that in order to study helical symmetry with this Hamiltonian, it is necessary to have $t_3\gg t_1$ and $t_4\gg t_2$.
Operator $a_{lkn}$ ($a_{lkn}^+$) annihilates (creates) an electron at period l, ring k and site n. 
$n_N$ represents $n (mod\,N)$, $\delta_{ij}$ is the Kronecker delta, $h.c.$ is hermitian conjugate and $\alpha$ and $\beta$ are the corresponding phase changes between nearest sites in a particular ring and between nearest rings along the toroid respectively. 
They can be expressed in terms of AB flux ($\Phi$) as
\begin{equation}
\alpha=2\pi\frac{\Phi(\alpha)}{N\Phi_0}\,\, , \,\,\, \beta=2\pi\frac{\Phi(\beta)}{LN\Phi_0} .\label{alphabeta}
\end{equation}

Eigenfunctions of the hamiltonian, $H=\sum_{l,k,n}\sum_{l',k',n'} h_{lkn,l'k'n'}a^+_{lkn}a_{l'k'n'}$ can be written as an expansion of $\psi=\sum_{l,k,n}C_{lkn}a^+_{lkn}|0>$ where $|0>$ denotes the vacuum state.
Expansion coefficicents $C_{lkn}$ satisfies $\sum_{l'k'n'}h_{lkn,l'k'n'}C_{l'k'n'}=E_{lkn}C_{lkn}$. 
According to Fig. \ref{helical1}, system exactly repeats itself after translation along the toroid by $N$ rings. 
Therefore Bloch theorem applies and it gives $C_{l+1,k,n}=e^{i\gamma}C_{l,k,n}$ where $\gamma=\frac{2\pi}{L}s$ with $s=0,1,...,L-1$.

Bloch theorem partly digonalizes matrix $H_{lkn,l'k'n'}$ by quantized values of $\gamma$ with a reduced Hamiltonian matrix elements given by
\begin{eqnarray}
&&H_{kn,k'n'}=\left[-t_1\delta_{kk'}\delta_{n',n_N+1}e^{i\alpha} \right. \nonumber \\
&-&(t_3-t_1)\delta_{kk'}\delta_{n',n_N+1}\delta_{k,n_N+1}e^{i\alpha}  \nonumber \\
&-&t_2\delta_{n,n'}\delta_{k',k_N+1}\left[ \delta_{kN}e^{i(\beta+\gamma)} + (1-\delta_{kN})e^{i\beta}\right] \nonumber \\
&-&(t_4-t_2)\delta_{n,n'} (1-\delta_{kN})\delta_{k',k_N+1}\delta_{kn}e^{i\beta} \nonumber \\
&-&(t_4-t_2)\delta_{n,n'}\left. \delta_{kN}\delta_{k',k_N+1}\delta_{kn}e^{i(\beta+\gamma)} + h.c.\right] . 
\end{eqnarray}

Matrix $H_{kn,k'n'}$ should be diagonalized numerically which suggests that instead of $k,n$, we introduce $r=N(k-1)+n$ with $r$ changing from 1 to $N^2$.
For a given value of $\alpha$ and $\beta$, Hamiltonian has total of $LN^2$ eigenvalues since $H_{kn,k'n'}(\gamma)$ should be diagonalized for each value of $\gamma=\frac{2\pi}{L}s,\,s=0,1,...,L-1$.
Finding these eigenvalues accomplishes unitary transformation of creation operators from $a^+_{lkn}$ to $a^+_{rr'}$ which ensures that states $\psi_{rr'}=a^+_{rr'}|0>$ are orthogonal to each other and $a^+_{rr'}$ are canonical Fermi operators, i.e. $[a^+_{rs},a_{r's'}]_+=\delta_{ss'}\delta_{rr'}$.

For a given number of electrons in toroid, $N_e$, we calculate minimal energy which is sum of lowest $N_e$ out of $LN^2$ eigenvalues and also calculate total persistent currents both along the toroid and along the rings by variation of free energy of the system with the magnetic flux.

\begin{figure} [ht]
\centerline{\scalebox{0.45}{\includegraphics{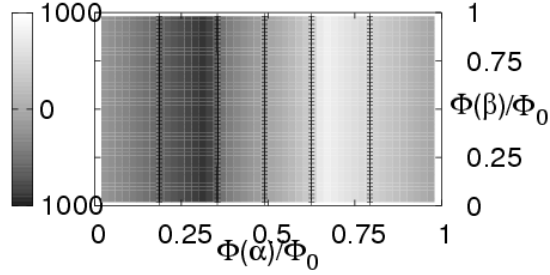}}}
\caption{\label{fhm-5} Contour plot of $I_c(\alpha)$ vs. Flux in the first helical model. Parameters are $t_1=t_3=10$, $t_2=t_4=1$, $N=L=10$ and $N_e=500$ (half filling).
}
\end{figure}

\begin{figure} [ht]
\centerline{\scalebox{0.45}{\includegraphics{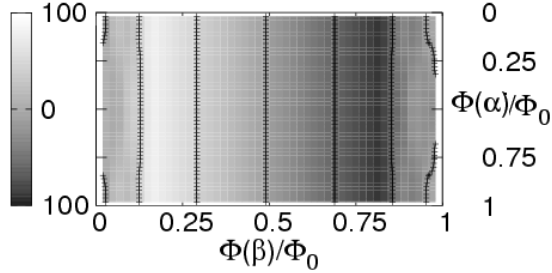}}}
\caption{\label{fhm-8} Contour plot of $I_l(\beta)$ vs. Flux in the first helical model. Parameters are $t_1=t_3=1$, $t_2=t_4=100$, $N=L=10$ and $N_e=500$ (half filling).
}
\end{figure}

Total persistent currents along the rings (circular) and the toroid (longitudinal) are perpendicular to each other.
In Fig. \ref{fhm-5}, we choose hopping parameters such that probability of finding electrons inside the rings is much more than along the toroid.
In this limit, $I_c(\alpha)$ dominates $I_l(\beta)\rightarrow 0$ and the coupling between these currents is very small (negligible) as shown in the figure.
In the opposite limit, where electrons have more probability of being along the toroid than inside the rings, $I_l(\beta)$ dominates $I_l(\alpha)\rightarrow 0$ and the coupling is also very small as shown in Fig. \ref{fhm-8}.

\begin{figure} [ht]
\centerline{\scalebox{0.45}{\includegraphics{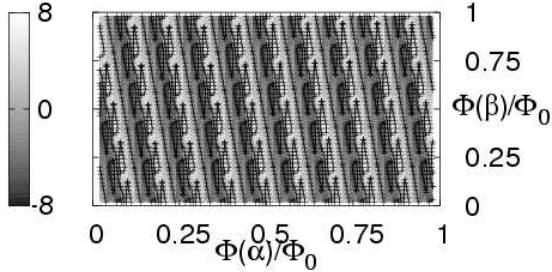}}}
\caption{\label{fhm-6a} Contour plot of $I_c(\alpha)$ vs. Flux in the first helical model. Parameters are $t_1=t_2=1$, $t_3=t_4=10$, $N=L=10$ and $N_e=500$ (half filling).
}
\end{figure}

\begin{figure} [ht]
\centerline{\scalebox{0.45}{\includegraphics{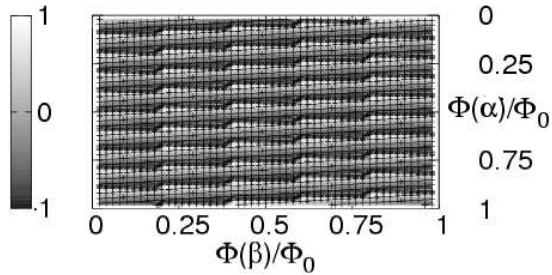}}}
\caption{\label{fhm-6b} Contour plot of $I_l(\beta)$ vs. Flux in the first helical model. Parameters are $t_1=t_2=1$, $t_3=t_4=10$, $N=L=10$ and $N_e=500$ (half filling).
}
\end{figure}

In order to understand the mixing of both symmetries, we choose another set of parameters such that probability of finding electrons along the helical path (Fig. \ref{helical1}) is much more than finding it elsewhere in the toroid.
This case is shown in Fig. \ref{fhm-6a} and \ref{fhm-6b}.
As opposed to the previous limits, we find that mixing symmetries has cross-effect on the system and both currents coupled to each other.
However, both currents in different directions are periodic in $\Phi$ with a period of $\Phi_0$ as expected in all cases.

\begin{figure}  [ht]
\centerline{\scalebox{0.4}{\includegraphics{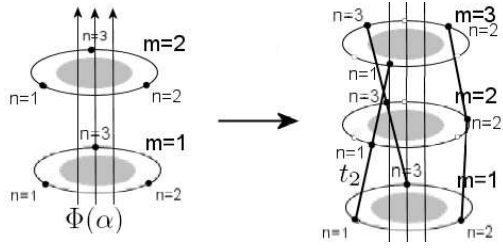}}}
\caption{\label{helical2}
         Left: Connected rings in the first model with $\Phi(\alpha)$ applied at the center.
         Right: Rotated rings by angle $2\pi s/LN$ for $s=1,N=3$ and $L=3$.}
\end{figure}

\emph{Second Helical Model:}
In order to satisfy necessary boundary conditions for the geometry of the toroid ring $m=LN+1$ should coincide with ring $m=1$. 
As shown in Fig. \ref{helical2}, we fix the position of the first ring and rotate rest ($LN-1$) of them by $2\pi s/LN$ where $s=0,1,...,LN-1$.
Note that each value of parameter $s$ specifies different geometry.
In this model, we considered all possible circular and cross hoppings both inside and between the rings. 

In the tight-binding approximation, system Hamiltonian becomes
\begin{eqnarray}
H&=&\sum_{m=1}^{LN}\sum_{n=1}^N\left[-t_1 a_{mn}^+a_{m,n_N+1}e^{i\alpha} \right. \nonumber \\
&-&t_2\left. a_{mn}^+a_{m_{LN}+1,n}e^{i\beta}e^{i\frac{s\alpha}{L}} + h.c. \right]  \nonumber \\
\end{eqnarray}
where $t_1$ is the hopping amplitude between the nearest-neighbour sites in a ring and $t_2$ is the cross-hopping amplitude between the rings.
Operators $a_{mn}$ ($a_{mn}^+$) annihilates (creates) an electron at ring m and site n.
$\alpha$ and $\beta$ are the corresponding phase changes between sites in a particular ring and different rings in the toroid respectively.
They are given by Eq. \ref{alphabeta}.

Hamiltonian can be diagonalized by discrete Fourier transformation from $a_{mn}$ in site to $b_{qk}$ in momentum representation as
\begin{equation}
a_{mn}=\frac{1}{\sqrt LN}\sum_{k,q}b_{qk}e^{i(kn\Delta_1+qm\Delta_2)} ,\label{anm}
\end{equation}
where $k=\frac{2\pi}{N\Delta_1}n$ and $n=0,1,2,...,N-1$ and $q=\frac{2\pi}{LN\Delta_2}m$ and $m=0,1,2,...,LN-1$. 
In diagonal form, Hamiltonian becomes
\begin{eqnarray}
H&=&-2\sum_{q,k}b_{qk}^+b_{qk}\left\{t_1\cos{(k\Delta_1+\alpha)}\right. \nonumber \\
&+&\left.t_2\cos{\left(q\Delta_2+\beta+\frac{s\alpha}{L}\right)}\right\} .\label{secondHD}
\end{eqnarray}
Eigenvalues of this Hamiltonian are periodic in $\Phi$ with a period of $\Phi_0$ and they are given by
\begin{eqnarray}
\epsilon_{mn}(\Phi)=-2t_1\cos{\left[\frac{2\pi}{N}\left(n+\frac{\Phi(\alpha)}{\Phi_0}\right)\right]} \nonumber \\
-2t_2\cos{\left[\frac{2\pi}{LN}\left(m+\frac{\Phi(\beta)}{\Phi_0}+\frac{s\Phi(\alpha)}{\Phi_0}\right)\right]} .\label{secondE}
\end{eqnarray}

Corresponding total persistent currents along the rings and the toroid (circular $\alpha$ and longitudinal $\beta$ currents) which are periodic in $\Phi$ with a period of $\Phi_0$ are perpendicular to each other and they are given by $I_{c}(\Phi(\alpha))=\sum_{m,n}\frac{d\epsilon_{mn}(\Phi)}{d\Phi(\alpha)}$ where summation is over number of electrons, $N_e$.
Ignoring the spin of electrons, for each value of flux we have
\begin{eqnarray}
I_{c}(\alpha)=\sum_{m,n}\left\{I_1\sin{\left[\frac{2\pi}{N}\left(n+\frac{\Phi(\alpha)}{\Phi_0}\right)\right]}\right. \nonumber \\
+\left.I_2\sin{\left[\frac{2\pi}{LN}\left(m+\frac{\Phi(\beta)}{\Phi_0}+\frac{s\Phi(\alpha)}{\Phi_0}\right)\right]}\right\} ,\label{secondIalpha}
\end{eqnarray}
\begin{eqnarray}
I_{l}(\beta)=I_2\sum_{m,n}\sin{\left[\frac{2\pi}{LN}\left(m+\frac{\Phi(\beta)}{\Phi_0}+\frac{s\Phi(\alpha)}{\Phi_0}\right)\right]} \label{secondIbeta}
\end{eqnarray}
where $I_1=-4c\pi t_1/N\Phi_0$ and $I_2=-4c\pi t_2s/LN\Phi_0$ are the current amplitudes.

\emph{Discussion and Conclusion:}
The geometric structure determines the electronic structure and thus the characteristics of the persistent current oscillations.
In this paper, we study symmetry mixing and cross-effects in a toroidal system which is threaded by magnetic flux both along ($\alpha$) and inside ($\beta$) the structure.
We consider a set of connected-mesoscopic rings and model helical symmetry by restricting hopping directions in our models.
The electronic structure calculated from the tight-binding model is given in Eq. \ref{secondE} for second helical model, however, we can not solve it for the first helical model and instead evaluate it numerically.
Since magnetic flux $\Phi(\alpha)$ and $\Phi(\beta)$ are in perpendicular directions, circular and longitudinal currents also flows in perpendicular directions.

In both models we consider electron transport inside the rings, however, we propose two models with hoppings in different directions between the rings.
In the first model, we allow only vertical hoppings between the neighbouring rings.
Since the system is periodic along the toroid, we include the Bloch condition and solve the final Hamiltonian matrix for energy eigenvalues and persistent currents numerically. 
Our results show that mixing perpendicular magnetix fluxes couples perpendicular currents with each other and both circular and longitudinal currents are periodic in $\Phi$ with a period $\Phi_0$.

In the second model, we consider electron transport with cross hoppings between the rings.
This coupling between perpendicular $\alpha$ and $\beta$ magnetic fluxes yields an extra component to the total circular current (\ref{secondIalpha}) with a period $\Phi_0/s$.
Since $s$ is a positive integer, total circular persistent currents have period $\Phi_0$ as expected.
In the special case, for $s=0$, all cross-hoppings are indeed now vertical hoppings and we recover the result of first model together with circular currents which have period $\Phi_0$.
We also note that Eq. \ref{secondIalpha} and Eq. \ref{secondIbeta} are in agreement with Eq. \ref{ringI} in the limits when $N\rightarrow 1$,$s=0$ and $t_2\rightarrow 0$.


Extra component (coupling) of circular persistent currents appears in both models.
These currents are vanishingly small in the limit of large number of rings, $L\rightarrow \infty$, as expected.
Note that, extra circular current component is due only to $\Phi(\alpha)$ and the presence of $\Phi(\beta)$ results only in longitudinal persistent currents along the toroid.
Our both model results are also in agreement with Lin \cite{c5} et al.. 
They showed that perpendicular $\Phi(\beta)$ through the carbon nanotube toroidal structures results in persistent current oscillations with a period $\Phi_0$.
To conclude, our calculations suggest that circular persistent currents in structures with helical symmetry have two components with periods $\Phi_0$ and $\Phi_0/s$.
However, total circular persistent current oscillations have $\Phi_0$ period.

\end{document}